\nonstopmode

\documentclass[aps,prb,twocolumn,showpacs]{revtex4}
\usepackage{graphicx}

\begin{document}
\title{Large contrast enhancement of graphene monolayers by angle detection}
\author{V. Yu}
\author{M. Hilke}
\affiliation{Department of Physics, McGill University, Montr\'eal, Canada H3A 2T8}
\date{\today}

\begin{abstract}
Exfoliated graphene monolayers are identified by optical inspection. In order to improve the monolayer detection, we investigate the angle dependence of the optical contrast of graphene on a 90nm SiO$_2$/Si substrate. We observe a significant enhancement of the visibility of graphene by changing the polarization and the angle of optical incidence. This method can be used to detect graphene on new substrate designs such as GaAs/AlAs based materials, which have a much cleaner surface.
\end{abstract}
\maketitle

Since its discovery in 2004 \cite{novo04}, graphene, a monolayer of carbon atoms packed into a honeycomb lattice led to an explosion of work. The high electronic mobility of intrinsic graphene makes it a very attractive candidate for electronics\cite{geim07}. Indeed, graphene on SiO$_2$ substrates have shown very high  mobilities \cite{tan07} at room temperature, but when the devices are cooled, mobilities do not improve significantly \cite{chen08}, which is in stark contrast to the highest mobilities in GaAs/AlGaAs \cite{pfeiffer03}. At low temperatures, the mobility is limited by scattering due to charged impurities \cite{chen08,hwang07} and the presence of ripples in graphene on SiO$_2$ \cite{ishig07,chen08-2,chen09}.

Graphene monolayers can be obtained by exfoliation or by epitaxy. However, in graphene epitaxy the quantum Hall effect is not clearly seen because of the coupling to the substrate \cite{didier08}, which leads to a decreasing mobility as a function of magnetic field. In contrast, uncoupled single layers of graphene can be deposited via micro\-mechanical cleavage of graphite flakes \cite{novo05}. Since the active electronic region in graphene is only one atom away from the substrate, the quality of the substrate will substantially influence the electronic properties as demonstrated by the increase of mobility for suspended graphene \cite{bol08}. However, it is very difficult to make advanced devices out of suspended materials. This brings us to consider the possibility to deposit graphene on different substrates, while still providing for a good visibility.

Graphene layers can be localized on oxidized Si wafers through changes of the reflection at optical wavelengths. Optimal thicknesses of the oxide layer are found to be 90 and 300 nm and lead to an optical contrast of about 15\% for a graphene monolayer \cite{blake07,aberg07,rodd07,chang07,jung07}. Adding a resist layer can enhance the contrast \cite{gteo08} but requires the deposition of a very uniform resist before localizing the interesting flakes. Here we discuss an alternate route to increase the contrast of a graphene monolayer. The main idea is to increase the contrast by polarizing and directing the incoming light at various angles.

\begin{figure}[!h]
	\centering
		\includegraphics[scale=0.7]{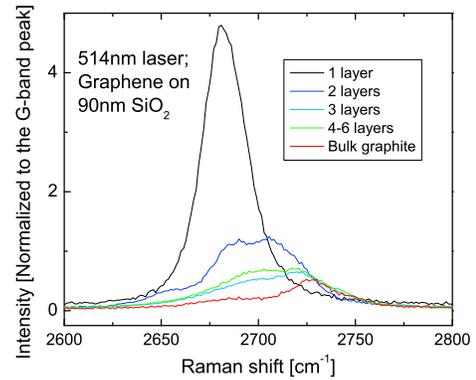}
		\vspace{-0.3cm}
		\caption{Raman spectrum for each individual layers. The 2D band peaks are normalized to the G band.}
	\label{Raman}
\end{figure}

Graphene is deposited using natural graphite from Asbury Carbon via mechanical exfoliation. In order to identify single layered graphene flakes, we performed calibration measurements using Raman spectroscopy as seen in Figure \ref{Raman}. This technique can be used as a reference for optical contrast coding \cite{gupta06} and maps the number of la\-yers to an optical contrast value. Once single layered graphene flakes were identified we measured the angle dependence of the contrast for different colors. We used a white light source and recorded the signal at different wavelengths.

\vspace{-0.2cm}
\begin{figure}[!h]
	\centering
		\includegraphics[scale=0.2]{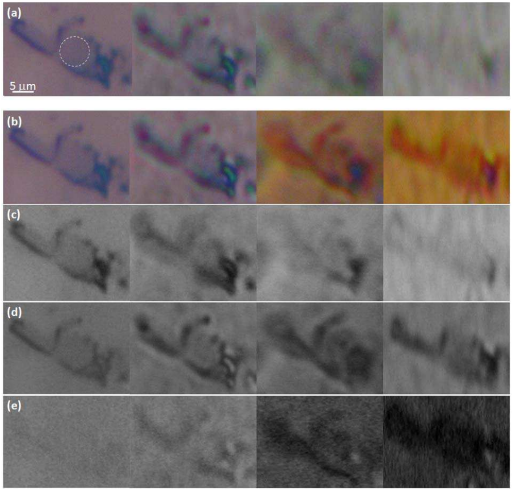}
		\caption{Thin graphene flakes on 90nm SiO$_2$ imaged on an optical microscope under (a) TM and (b) TE polarizations. (c) red, (d) green, and (e) blue images were obtained by selecting the corresponding wavelengths. Each column represents images taken at 0$^\circ$, 23$^\circ$, 55$^\circ$, and 60$^\circ$ respectively. The dotted circle is the spot used for the contrast measured of a monolayer.}
		\vspace{-0.3cm}
	\label{fig:contrast_pics}
\end{figure}

\vspace{-0.15cm}
The first column of Figure \ref{fig:contrast_pics} is a typical image of a graphene mono-layer at normal incidence. By increasing this angle, corresponding to images from left to right, the colors of the SiO$_2$ and graphene change. Figure \ref{fig:contrast_pics}(a) is obtained with TM (Transverse Magnetic) polari\-zation and as the angle increases, a decrease in contrast of graphene is observed for all wavelengths. With TE (Transverse Electric) polari\-zation, Figure \ref{fig:contrast_pics}(b) shows a change in color with increased angle. In order to analyze the contrast, we selected a red, green and blue wavelength respectively (Figure \ref{fig:contrast_pics}(c)(d)(e)). A decrease of contrast is seen in the red as the angle increases, while the contrast for green and blue increases. We can understand these changes in terms of the following simple multilayered optical model.

\begin{figure}[!h]
	\centering
\begin{tabular}{c}\includegraphics[scale=0.25]{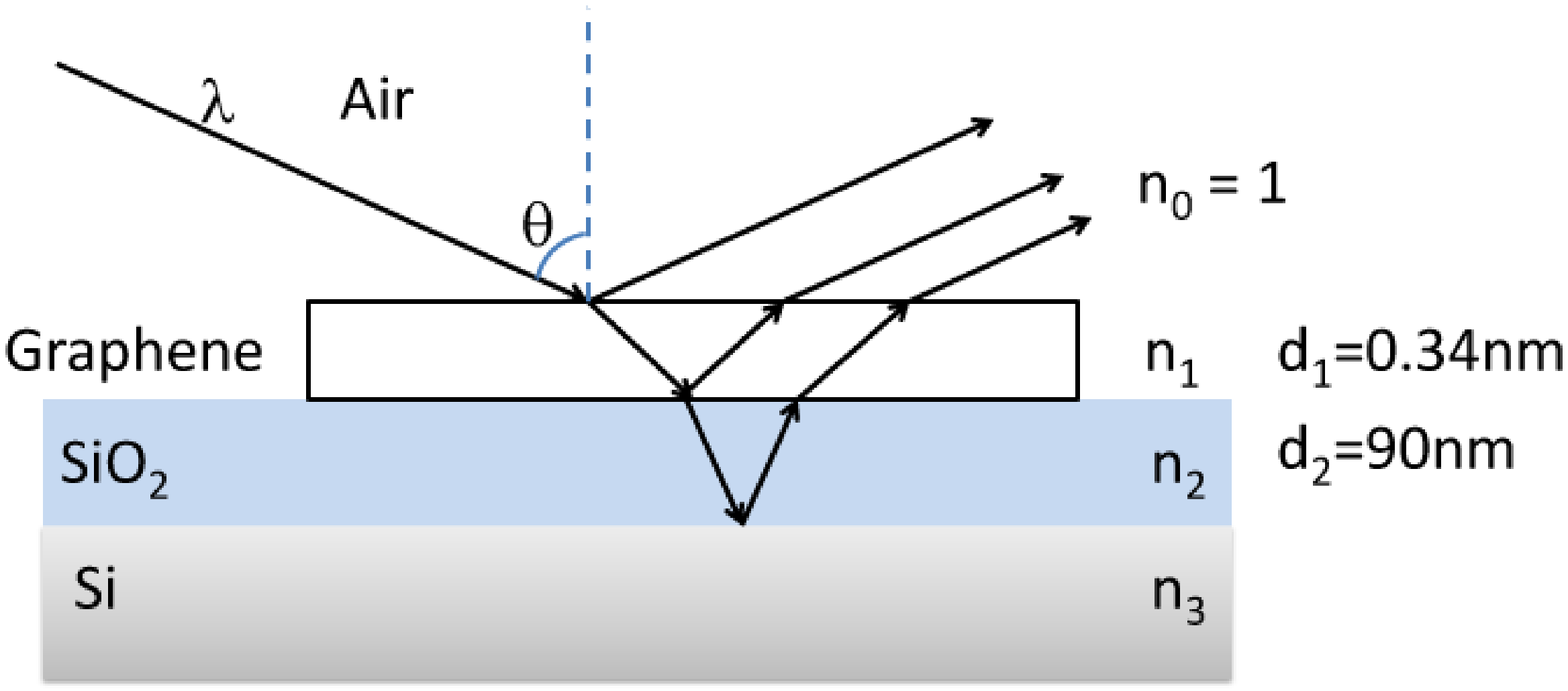} \\ \includegraphics[scale=0.2]{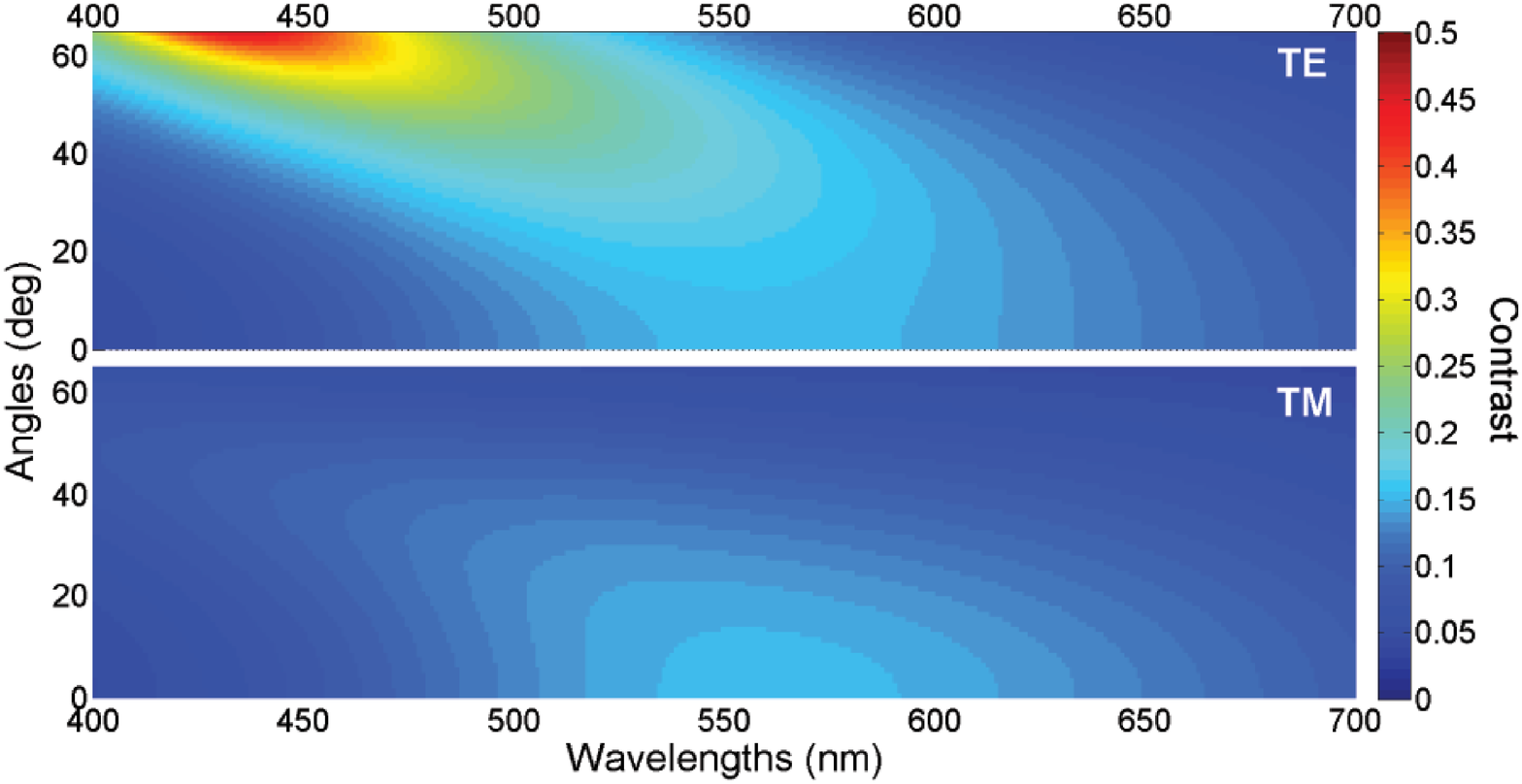}\end{tabular}
		\caption{Top: Model used for our analysis. Optical reflection for a 3-layer system. Bottom: Contrast plot for TE and TM polarizations as function of wavelength and angles for a 90nm SiO$_2$/Si substrate. The expected contrast is shown on the color scale found on the right.}
	\label{fig:setup}
\end{figure}

 We consider a SiO$_2$/Si substrate with an oxide thickness of 90nm (taken from the measured value of 90$\pm$2nm), while the thickness of a graphene monolayer is set to 0.34nm. We then consider an incoming light beam of wavelength $\lambda$ (the external light source), which hits a trilayer structure (graphene/SiO$_2$/Si) and passes through a series of reflections and transmissions as seen in Figure \ref{fig:setup}. The Si layer is assumed to be semi-infinite. The index of refraction of SiO$_2$ (n$_2$) and Si (n$_3$) are wavelength dependent and their values are given in Ref. \onlinecite{handb91} and \onlinecite{henrie04}. The index of refraction of graphene is assumed to be the same as that of bulk graphite (n$_1$ $\simeq$ 2.6 + 1.3\textit{i}) \cite{handb91}.

The results are obtained by considering a system of 3 layers (Figure \ref{fig:setup}) in which each layer j can be re\-presented by a transfer matrix. For a TM polarization, the magnetic field component H$^{z}$ is parallel to the layers and the electrical component E$^{y}$ is at the incident angle $\theta$. The general form of the transfer matrix for a layer j can be written as \cite{hilke08,chilwell84}:

\begin{equation}
\left(\begin{array}{c}
H^z_{j-1}\\E^y_{j-1}\end{array}\right)=\underbrace{\left(\begin{array}{cc} \cos(\phi_j) &
-i\sin(\phi_j)/\gamma_j\\-i\sin(\phi_j)\gamma_j & \cos(\phi_j)
\end{array}\right)}_{M_j}\left(\begin{array}{c}
H^z_{j}\\E^y_{j}\end{array}\right). \label{iter1}
\end{equation}

The relation of the field components for a wavelength $\lambda$ can be expressed as the product of transfer matrices (M = $\Pi^{2}_{j=1}$M$_j$) with the reflectance given by

\begin{equation}
R = \bigg|\frac{M_{11} + \gamma_3 M_{12} - M_{21}/\gamma_0 - \gamma_3M_{22}/\gamma_0}{ M_{11} + \gamma_3 M_{12} + M_{21}/\gamma_0 + \gamma_3M_{22}/\gamma_0}\bigg|^2,
\label{reflect}
\end{equation}
where M$_{kl}$ are the matrix elements of M. The matrix M$_j$ in Equation \ref{iter1} depends on the index of refraction ($n_j$) and width ($d_j$) of the corresponding layer, with $\phi_j=(2\pi d_j/\lambda)\sqrt{n_j^2-\sin^2(\theta)}$, $\gamma_j=(z_0/n_j^2)\sqrt{n_j^2-\sin^2(\theta)}$, and the vacuum impedance $z_0$. For the TE polarization, the expressions have to be substituted by $H^z\rightarrow E^z$, $E^y\rightarrow -H^y$, and $\gamma_j=z_0^{-1}\sqrt{n_j^2-\sin^2(\theta)}$. Once the reflectance is computed, the contrast is given by $C=(R_1-R_2)/R_1$, where $R_1$ is the reflectance without graphene and $R_2$ the reflectance in the presence of graphene.

\begin{figure}[!h]
	\centering
		\includegraphics[scale=0.75]{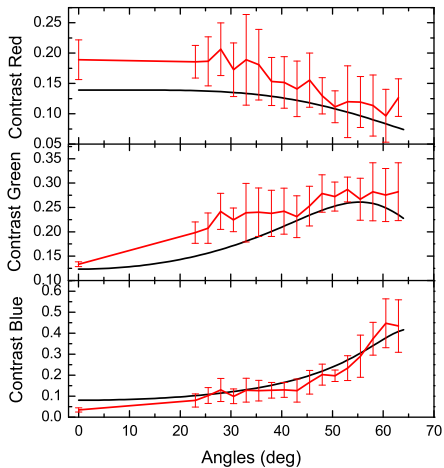}
		\vspace{-0.2cm}
		\caption{Comparison of experimental (red curve) and calculated (black curve) contrast plot as a function of angle at a fixed wavelength. (a) Red contrast plot at a wavelength of 620nm (b) Green contrast plot at 495nm. (c) Blue contrast plot at 450nm. The error bars are the standard deviations for 3 different data sets and the line is a guide to the eye.}
		\vspace{-0.3cm}
	\label{fig:graph_exp}
\end{figure}

While several groups have computed the contrast at normal incidence \cite{blake07,jung07,gteo08}, we show here the results as a function of incident angle for different polarizations. Figure \ref{fig:setup} shows the evolution of the contrast as function of angle in the visible spectrum for two polarizations. While no contrast enhancement is obtained for TM polarizations, TE polarizations show a significant enhancement in the blue wavelengths. We compare the calculated contrast with the data for different wavelengths in Figure \ref{fig:graph_exp}. The data follows the calculated one within the error bars obtained from the standard deviation of several different data sets. Clearly, the blue and green wavelengths show a significant increase in contrast as a function of angle, reaching even 40\% for blue.

\begin{figure}[!h]
	\centering
		\includegraphics[scale=0.25]{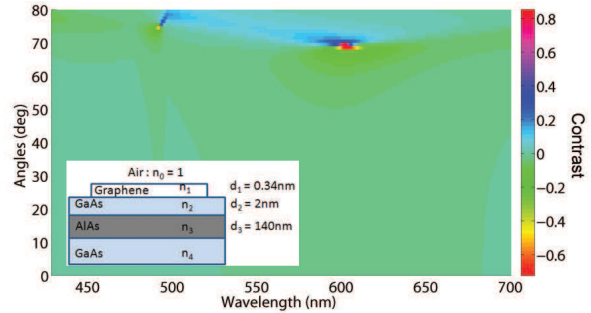}
		\caption{Contrast plot as a function of angles and wavelength with fixed thickness of 2 nm of GaAs, 140nm of AlAs and TM polarizations. The inset shows the structure used for our calculation.}
		\vspace{-0.1cm}
	\label{fig:AlAs_0to65}
\end{figure}

Based on our understanding of the contrast on SiO$_2$, we now turn to the possibility of enhancing the contrast of graphene on other substrates, which would offer a smoother and cleaner electrical surface than SiO$_2$. Here, we consider an optimal GaAs/AlAs/GaAs hetero\-structure which can be grown by molecular beam epitaxy.  It was previously reported in Ref. \onlinecite{gteo08} that the optimal contrast for a homostructure of GaAs is -0.0059. Here, we propose a 4-layer Fabry-P\'erot structure suitable to increase the contrast for the detection of a single layer graphene as shown in the inset of Figure \ref{fig:AlAs_0to65}. For this structure, the intensity of the reflected waves is calculated by considering an additional layer, leading to an extra transfer matrix in Equation 1. The first GaAs layer is described by a thickness of d$_2$ and its refractive index is wavelength dependent \cite{handb91}. For simplicity, the thickness d$_2$ is fixed at 2nm since the smallest layer of GaAs improves the overall contrast of graphene. It is important to have GaAs as a cap layer since it is a much cleaner surface than AlAs, which oxidizes much more. Next, the thickness of the AlAs layer ($d_3$) was varied and found to be optimal at 140nm. In Figure \ref{fig:AlAs_0to65} we show the contrast as a function of incident angle. At normal incidence, a contrast of 2\% was obtained, which improves the contrast by almost a factor of 4 compared to a monostructure of GaAs. As the angle increases to 65 degrees, the expected contrast is approximately 14\%. This contrast is identical to graphene on SiO$_2$ at normal incidence and should allow the identification of graphene monolayers on GaAs. Interestingly, the contrast for GaAs is only enhanced for a TM polarization as opposed to a TE polarization in the SiO$_2$ case.

In summary, we found a simple way to increase the contrast of a single layer of graphene on SiO$_2$ by simply tilting the sample and the light source. This method can then be applied to other substrates, such as GaAs/AlAs heterostructures, which have much cleaner surfaces in order to substantially increase the mobilities.

We would like to acknowledge helpful discussions with T. Szkopek, and support from NSERC, FQRNT, and RQMP.

\end{document}